\documentclass[aps,prb,showpacs,12pt]{revtex4}
\usepackage{amsmath,amssymb}
\usepackage{graphicx,latexsym}
\def\CBT{CsBi$_4$Te$_6$}
\def\beq{\nopagebreak \begin{equation}}
\def\eeq{\end{equation}}
\def\bmpt#1{\begin{minipage}[t]{#1\linewidth} \vspace{0pt}}
\def\emp{\end{minipage}}
\def\bk{{\bf k}}

\def\df{\Bigl[-\frac{\partial f_\mu(T;\varepsilon)}{\partial \varepsilon} \Bigr]}

\begin{document}
\title{Electronic structure and transport in CsBi$_4$Te$_6$.}
\author{Lars Lykke, Bo B. Iversen, Georg K. H. Madsen}
\affiliation{Dept. of Chemistry, University of Aarhus, DK-8000 {\AA}rhus C, Denmark}
\email{georg@chem.au.dk}
\date{\today}
\begin{abstract}
The band structure of the novel low-temperature thermoelectric material, \CBT, is calculated and analyzed using the semi-classic transport equations. It is shown that to obtain a quantitative agreement with measured transport properties a band gap of 0.08~eV must be enforced. A gap in reasonable agreement with experiment was obtained using the generalized gradient functional of Engel and Vosko. We found that the experimental $p$-type sample has a carrier concentration close to optimal. Furthermore the conduction bands have a form equally well suited for thermoelectric properties and we predict that an optimally doped $n$-type compound could have thermoelectric properties exceeding those of the $p$-type.
\end{abstract}
\maketitle
\section{Introduction}
The search for new thermoelectric materials is a quest to maximize the
dimensionless figure of merit $zT=(\sigma T/\kappa) S^2$, where $S$ is
the Seebeck coefficient and $\sigma$ and $\kappa$ are the electronic
and thermal conductivities respectively. $zT$ quantifies the
performance of a thermoelectric and one must therefore maximize the
power-factor $S^2\sigma$ and minimize $\kappa$. As $S$, $\sigma$ and
$\kappa$ are coupled and all depend strongly on the detailed
electronic structure, carrier concentration and crystal structure, finding new
compounds with large values of $zT$ is a difficult task.
Still, several completely new types of materials, with complex
crystal structures and $zT$'s exceeding the presently used alloys, have recently been found.\cite{CsBiTe,TlBiTe,AgPbSbTe,znsb_natmat} 

\CBT\ is remarkable because of its thermoelectric properties at low
temperatures ($zT\approx 0.8$ at 225 K).\cite{CsBiTe} It has a complex layered
crystal structure and some intriguing direct Bi$-$Bi bonds.\cite{CsBiTe} It would
therefore be interesting if a direct link between the bonding and band
structure of CsBi$_4$Te$_6$ and the thermoelectric properties could be
established, as has been done earlier for other thermoelectric
compounds.\cite{davidskut2,gmeugage,sofote,bertini04,chaput05} One
study of the band structure of CsBi$_4$Te$_6$ has been
published\cite{CsBiTe_larson} and very recently this study was used to
interpret the results of a systematic experimental study of \CBT\
doping.\cite{CsBiTe2} Though the earlier band structure study did report the
effective masses\cite{CsBiTe_larson}, we wish to improve the
quantitative link between the band structure and the thermoelectric
quantities. Furthermore, despite using the same LAPW based method, we obtain a somewhat different band structure than was found earlier.\cite{CsBiTe_larson} We suspect that this is due to relativistic
effects being poorly treated in the earlier study.\cite{CsBiTe_larson}
These are extremely important due to the large spin-orbit splitting of
the Bi-$6p$ valence states. The challenge of obtaining a correct band
structure for the bismuth-telluride compounds is probably best
illustrated by the attempts to calculate the band structure of the well-known thermoelectric compound Bi$_2$Te$_3$. De-Hass-van Alphen
experiments\cite{rayne} and angle-resolved photo-emission
studies\cite{greanya} have shown that both the lowest conduction band
(LCB) and the highest valence band (HVB) have six-fold degenerate band edges.
Only recently has this been reproduced by band structure
calculations\cite{freemanbite1,BiTep12,sofote} and only very recently
has also the gap been correctly predicted.\cite{freemanbite2}

The paper is organized as follows: First we briefly review the crystal structure and measured thermoelectric properties. We then discuss the size of the band gap and compare the theoretical prediction with the experimental measurements. The dependency of the transport properties on the carrier concentration is then reported and finally we discuss the band structure and bonding.

\section{Structure and measured transport}
\begin{figure}
\caption{(color online) Structure of \CBT. The structure consists of long parallel rows of BiTe$_6$ octahedra (green/light grey) stitched together by Te$_5$Bi$-$BiTe$_5$ bonds (blue/dark square based pyramids and bonds). The BiTe-layers are separated by layers of Cs atoms which act as electron donors to the framework. The unit cell shown is the $C2/m$ monoclinic cell $a=51.9205$~\AA, $b=4.4025$~\AA, $c=14.5118$~\AA\ and $\beta=101.4800^\circ$. 
}
\label{fig:struct}
\end{figure}
The structure of \CBT\ is shown in Fig.~\ref{fig:struct}. The atoms are arranged in Cs layers and Bi$_4$Te$_6$ slabs with Bi$-$Bi bonds stitching the slabs together along the $a$-axis. The Bi$-$Bi bond is unusual as it involves a reduction of Bi$^{3+}$ to Bi$^{2+}$ seldom seen in Bi-chalcogenide systems.\cite{CsBiTe_larson} The inner located Bi atoms are coordinated to six Te atoms in a distorted octahedral whereas Bi atoms involved in Bi$-$Bi bonding are coordinated to five Te atoms.

\CBT\ has a side centered monoclinic unit cell. In the $C$-centered setting the $a$-axis is considerably longer than the $b$- and $c$-axes, Fig.~\ref{fig:struct}. The slabs stretch out infinitely along the $b$-axis making \CBT\ a 1D-like structure along the short $b$-axis. Von Neumanns principle relates the point group symmetry of the crystal structure (monoclinic) to the tensor properties of the crystal. The monoclinic symmetry with $\beta\neq90^\circ$ means that the conductivity tensor will have the symmetry
\beq 
\boldsymbol{\sigma}=\begin{pmatrix}
  \sigma_{aa} & 0 & \sigma_{ac} \\
  0      & \sigma_{bb} & 0 \\
  \sigma_{ca} & 0 & \sigma_{cc} \\
\end{pmatrix}
\eeq 
and similarly for the Seebeck coefficient. 

Experimentally a good conductivity was only found parallel to the short $b$-axis,\cite{CsBiTe,CsBiTe2} as would be expected from the crystal structure. Consequently good thermoelectric properties were only found when measured parallel to the $b$-axis.\cite{CsBiTe,CsBiTe2} The measured Seebeck coefficient rises monotonically up to approximately 275~K ($S\approx 175 \mu$V/K) and starts to decrease at higher temperatures. This is in good agreement with the very narrow band gaps, ranging from 0.04 to 0.1~eV, found experimentally\cite{CsBiTe,greanya,CsBiTe2} and means that \CBT\ only has favorable thermoelectric properties at low temperatures. It is also important for the thermoelectric properties of \CBT\ that a high electrical conductivity ($\approx 1450$~S/cm at 250~K) and an extremely low lattice thermal conductivity of $\kappa_l=0.6$~W/mK (calculated by using the Wiedemann-Franz law to subtract $\kappa_e$ from the measured $\kappa$) parallel to the $b$-axis were found.\cite{CsBiTe,CsBiTe2}

\section{Computational method}
\label{sec:comp}
\subsection{Electronic structure}
The calculations were performed using the L/APW+lo method\cite{gmapwlo} as implemented in the WIEN2k code.\cite{wien2k} A plane wave cutoff defined by min$(R_{\alpha})$max$(k_n)=6$ and sphere sizes of 2.7 were used. The exchange-correlation potential was calculated with the Perdew-Burke-Ernzerhof (PBE)\cite{pbe} and the Engel-Vosko (EV)\cite{ev} generalized gradient approximations (GGAs). 36 $k$-points on a shifted mesh in the IBZ were used for the SCF calculations

As will be discussed later we find a band gap of zero using the PBE-GGA. This disagrees with the earlier band structure study of \CBT, which found a gap of 0.04~eV.\cite{CsBiTe_larson} A gap of 0.04~eV is in reasonable agreement with experiment, but we believe this was caused by a cancellation of errors. We can see two possible sources of error in the previous paper:  either a poor treatment of spin-orbit coupling or an inadequate $k$-point sampling. 

In WIEN2k spin orbit coupling is included in a second variational step.\cite{kunesurhal} The size of the second variational basis set is controlled by an energy cut-off which limits the number of eigen vectors used. Furthermore $p_{1/2}$ local orbitals can be included in the second variational step.\cite{p12} Because of a strong spin-orbit splitting of the Bi $6p$-states at the Fermi-level, a correct description of relativistic effects is extremely important and the second variational basis-set must be well converged. We first performed a calculation using the default energy cut-off (1.5~Ry) of the WIEN code and did find that this induces a small gap of 0.012~eV. To test convergence we then performed two calculations: one setting a high energy cut-off (6.0~Ry) and another with a smaller cut-off (2.0~Ry) but including $p_{1/2}$ local orbitals\cite{p12} for the Bi atoms at energies close to the Fermi level. Both calculations found a band gap of zero and a very good agreement between the two calculations was found. No details about the second variational basis set were reported in the earlier paper,\cite{CsBiTe_larson} but it should be pointed out that a published study on Bi$_2$Te$_3$,\cite{BiTe_larson} had to be corrected due poor treatment of the $p_{1/2}$ contribution.\cite{BiTep12}

The earlier paper\cite{CsBiTe_larson} found the LCB minimum at the (0.881,0.881,0.175) position in the primitive reciprocal basis used in that study. This is the (0.238,0,-0.175) position in the $C$-side reciprocal basis used here. The earlier minimum position is not in complete agreement with the minimum position found here: (0.259,0,-0,166). The minimum of the LCB calculated band gap is not at a high symmetry position which means that an adequately dense $k$-mesh is very important for the exact location. The earlier paper does not report the procedure used to find the minimum, but only reports using 13 $\bk$ points in the IBZ for the self-consistent calculation, which is certainly to coarse a mesh for an exact location of the LCB minimum. We determined the minimum on the 1330 $\bk$ point mesh used in the transport calculations discussed below.

\subsection{Transport}

\begin{figure}[t]
\center
\includegraphics[width=.5\linewidth]{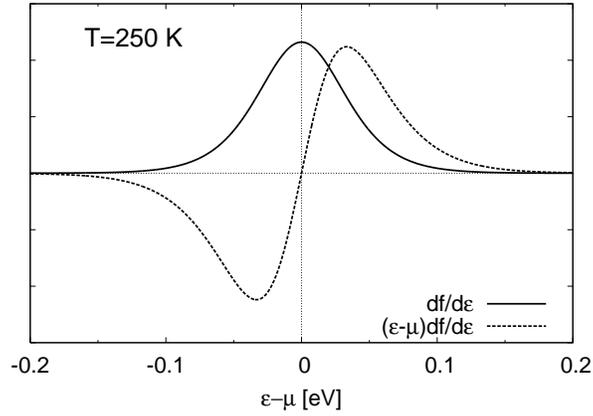} 
\caption{Integrand factors $\partial f/\partial \varepsilon$ and $(\varepsilon-\mu)\partial f/\partial\varepsilon$ in Eqs.~(\ref{eq:L1})-(\ref{eq:L2}) at $T=250$~K. Arbitrary $y$ units.}
\label{fig:dfde}
\end{figure}
The band structure was analyzed using semi-classic Boltzmann theory\cite{allenrev} and the rigid band approach. The rigid band approach to conductivity is based on the transport distribution
\begin{equation}
\sigma_{\alpha\beta}(\varepsilon)= \frac{1}{N}\sum_{i,{\bf k}} \sigma_{\alpha\beta}(i,{\bf k}) \frac{\delta(\varepsilon-\varepsilon_{i,{\bf k}})}{d\varepsilon}
\label{eq:transdist}
\end{equation}
The $\bk$-dependent transport tensor is given as
\beq
\sigma_{\alpha\beta}(i,{\bf k})=e^2\tau_{i,{\bf k}} v_\alpha(i,{\bf k}) v_\beta(i,{\bf k}) 
\label{eq:sigxx} 
\eeq
where $\tau$ is the relaxation time and $v_\alpha(i,{\bf k})$ is a component of the group velocities. The transport coefficients can be calculated as a function of temperature and chemical potential by integrating the transport distribution
\begin{gather}
\sigma_{\alpha\beta}(T;\mu)=\frac{1}{\Omega}\int\! \sigma_{\alpha\beta}(\varepsilon) \df d\varepsilon \label{eq:L1} \\
\nu_{\alpha\beta}(T;\mu)=\frac{1}{eT\Omega} \int\! \sigma_{\alpha\beta}(\varepsilon) (\varepsilon-\mu)\df d\varepsilon 
\label{eq:L2}
\end{gather}
The bands, and hence $\sigma(\varepsilon)$, are left fixed (thus ``the rigid band approach'') and therefore only one band structure calculation needs to be performed per compound. The number of carriers is changed by varying the chemical potential in Eqs.~(\ref{eq:L1}-\ref{eq:L2}). Fig.~\ref{fig:dfde} shows the integrand factors in Eqs.~(\ref{eq:L1}-\ref{eq:L2}) at $T=250$~K. It is seen that the distribution is quite broad and the transport coefficients are thus a sum over several Fermi surfaces. It is therefore very important that the band is correctly calculated.

In Eqs.~(\ref{eq:sigxx}) the relaxation time is unknown. Our approach in the present work is too treat it as a constant. The Seebeck coefficient, $S=\sigma^{-1}\nu$, is then independent of $\tau$ and can thus be calculated on an absolute scale. The conductivity can only be calculated with respect to the relaxation time. As $\kappa_l$ and $\tau$ are not readily available from band structure calculations they must somehow be included as parameters.

For the transport calculation, eigen energies at 1330 $k$-points on a non-shifted mesh in the IBZ were calculated. For calculation of the necessary derivatives, Eq.~(\ref{eq:sigxx}), the program BoltzTraP was used.\cite{BoltzTrap} BoltzTraP relies on a well tested smoothed Fourier interpolation to obtain an analytical expression of the bands.\cite{BoltzTrap} The original $k$-mesh was interpolated onto a mesh five times denser then the original. 

For computational reasons our calculations were carried out in the $B$-centered setting of the cell with $a=51.9205$~\AA, $b=51.0530$~\AA, $c=4.4025$~\AA\ and $\gamma=163.8256^\circ$, where the $c$-axis in the $B$-centered setting is parallel to the $b$-axis in the $C$-centered cell. However, we have converted all transport properties and band structures back into the basis of the $C$-centered cell and will discuss the transport properties parallel to the $b$-axis of the $C$-centered cell.

\section{Results and discussion}
\subsection{Band gap}
\begin{figure}
\bmpt{.02}
(a)
\emp
\bmpt{.47}
\includegraphics[width=.95\linewidth]{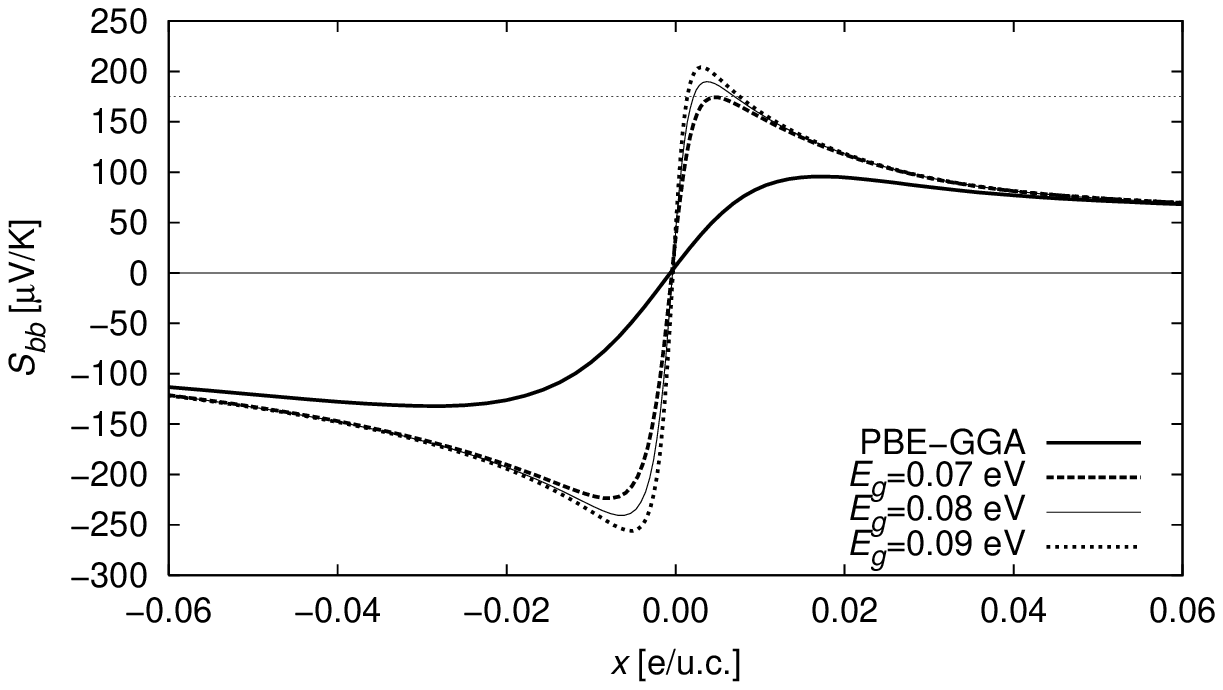}
\emp
\bmpt{.02}
(b)
\emp
\bmpt{.47}
\includegraphics[width=0.95\linewidth]{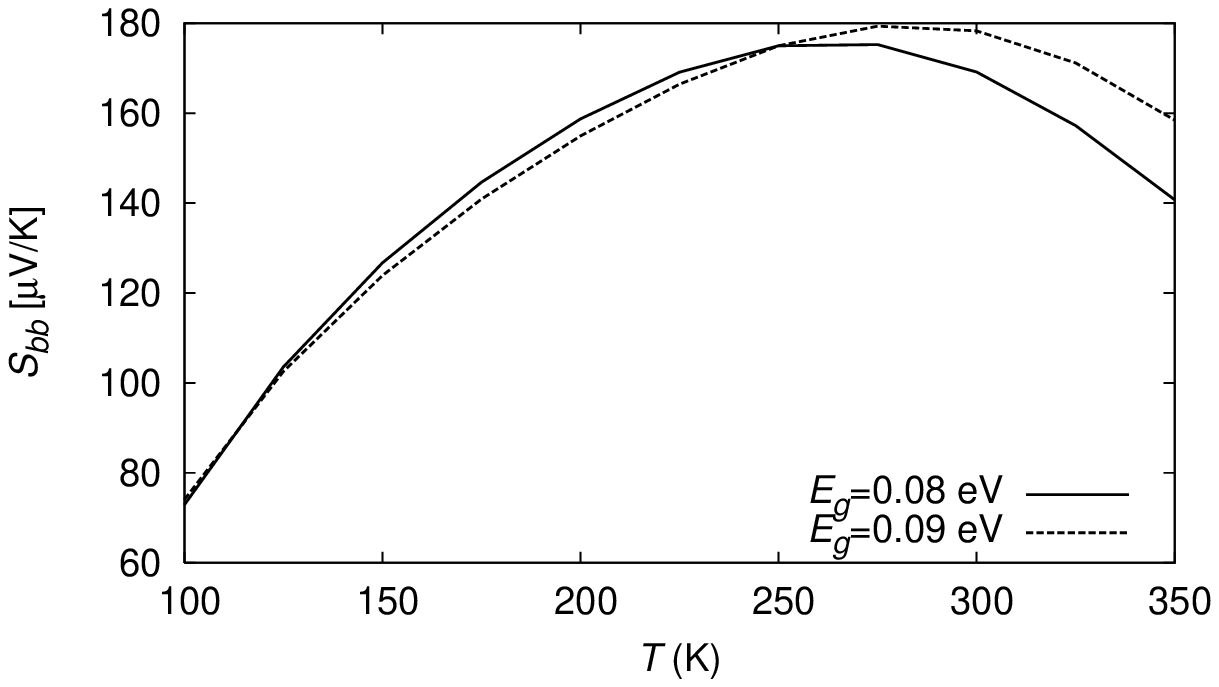}
\emp
\caption{ The $bb$ component of the Seebeck coefficient tensor at various fixed band gaps. (a) $S_{bb}$ vs carrier concentration at a fixed temperature of $T=250$~K. The dotted horizontal line marks the $S_{bb}=175$~$\mu$V/K found experimentally at 250~K. (b) $S_{bb}$ vs temperature. The carrier concentration for both curves has been fixed so that $S=175$~$\mu$V/K at 250~K. }
\label{fig:Seebeck}
\end{figure}

The calculated band structure of \CBT, which will be discussed in detail later, is predicted to have an indirect band gap of zero. This disagrees with the band gap of $0.04-0.1$~eV, found experimentally.\cite{CsBiTe,greanya,CsBiTe2} As a result of the underestimated band gap the calculated Seebeck coefficient of the standard PBE-GGA calculation is too low at elevated temperatures. This is illustrated in Fig.~\ref{fig:Seebeck}a, where the calculated $S$ (full line) does not agree with experiment, irrespective of carrier concentration.

The underestimation of the band gap is a well known problem of Kohn-Sham theory.\cite{gungap,rexgap,gwgap1} Experience suggests that the shape of the bands are correct and that most of the problem can be amended by a rigid shift of the conduction band. We therefore introduced a gap between the minimum energy of the LCB and the highest energy of the HVB by hand. The magnitude off the gap was then estimated by comparing the calculated Seebeck coefficient with the experimental values.\cite{CsBiTe} The first observation needed to fix the gap was that a gap larger than 0.06~eV is needed in order to obtain the experimental value of $S=175\ \mu$V/K at 250~K, Fig.~\ref{fig:Seebeck}a. Secondly, experiments have shown that the Seebeck coefficient starts to decrease at temperatures above approximately 250~K.\cite{CsBiTe} Fig.~\ref{fig:Seebeck}b shows that with a gap of 0.09~eV the Seebeck coefficient continues to increase up to approximately 300~K which gives an upper-bound of the gap. With a gap of 0.08~eV the experimental behavior is well reproduced and in good agreement with the band gap of 0.04-0.08~eV estimated from the experimental transport coefficients.\cite{CsBiTe2}

\subsection{The Engel-Vosko gap}
The problem of underestimating the band gap in DFT is often discussed and improvements have been suggested using the GW approximation\cite{rexgap,gwgap1,rexgap2,wkugw} or by including exact exchange.\cite{gksgap,exxgap1,freemanbite2,sharmagap} In some previous papers\cite{evtest,gmeugage} it has been suggested that one route to a better band gap could be the EV-GGA.\cite{ev} It is known to give very poor total energy differences, but for a number of narrow band gap semiconductors the calculated band gaps (i.e 1.23~eV for Si, 0.39~eV for Ge and 1.04~eV for GaAs) compare much better with experiment than the PBE-GGA. As a recent example we predicted a band gap of 0.4~eV for the type-VIII Eu$_8$Ga$_{16}$Ge$_{30}$ clathrate\cite{gmeugage} which was subsequently confirmed experimentally.\cite{VIIIgap}

We have applied the EV-GGA to \CBT, retaining the other computational parameters as reported in section~\ref{sec:comp}, resulting in $E_g=0.06$~eV, in good agreement with both experiment and the above analysis. To illustrate the usefulness of the EV-GGA we have also performed calculations on Bi$_2$Te$_3$. Using the PBE-GGA we obtain a band gap of 0.11~eV in agreement with an earlier calculation.\cite{sofote} Using the EV-GGA, but otherwise exactly the same computational parameters as earlier\cite{sofote}, we obtained a gap of 156~meV in excellent agreement with $E_g=154$~meV calculated using the sX-LDA\cite{freemanbite2} functional and in good agreement with the zero-temperature extrapolated experimental value of 162~meV.\cite{freemanbite2} 

The consistency of the better band gaps with the EV-GGA is thought provoking and could hint to an underlying reason. A possible explanation could lie in the construction of the EV-GGA which put used the virial relation is used to construct a functional which reproduces atomic exchange potentials better than the usual functionals.\cite{ev} It is known that the one-electron potential in the atomic limit jumps discontinuously at integer electron values\cite{atomlimit}, and that this discontinuity can give a large contribution to the band gap.\cite{gungap} It is possible that the EV-GGA, where , also gives a better reproduction of this discontinuity. However, the better band gaps of the EV-GGA are a phenomenological observation that should warrant further investigation.

\subsection{Optimal carrier concentration}

\begin{figure}
\bmpt{.02}
(a)
\emp
\bmpt{.47}
\includegraphics[width=0.95\linewidth]{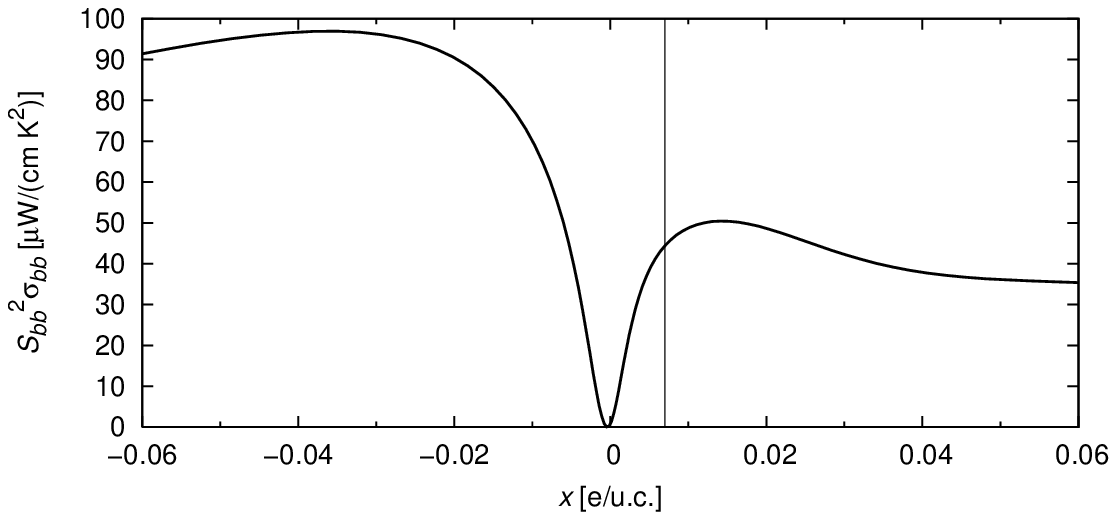}\\
\emp
\bmpt{.02}
(b)
\emp
\bmpt{.47}
\includegraphics[width=0.95\linewidth]{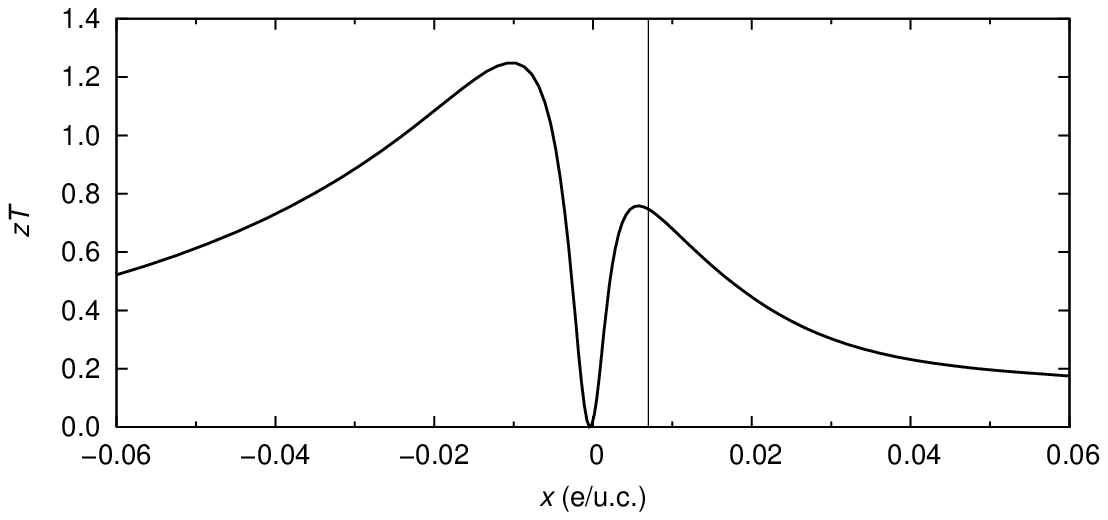}
\emp
\caption{(a) Power factor $S^2\sigma$ vs doping at 250~K. (b) Figure of merit ($zT$) vs doping at 250~K. The parameters $\tau=13.7\times 10^{-14}$~s and $\kappa_l=0.6$~W/mK were used (see text). $\kappa_e$ was calculated using the Wiedemann-Franz law. The vertical line indicates the carrier concentration that agrees with experiment.}
\label{fig:Power}
\end{figure}
Effective masses ($m^*$) can be calculated from the band structure as a function of carrier concentration and temperature.\cite{BoltzTrap} The effective masses are related to the conductivity through the relaxation time approximation: $\sigma=e^2(n/m^*)\tau$. From the band structure we obtain $e^2 n/m^*_{bb}=\sigma_{bb}/\tau=1.06\times10^{18}$~($\Omega$ m s)$^{-1}$ at the estimated experimental carrier concentration. Using the experimental conductivity of $\sigma_{bb}=1450$~S/cm at 250~K we obtain a relaxation time of $\tau=13.7\times 10^{-14}$~s.  Thus $\sigma_{bb}/\tau$ is not unusually high but $\tau$ is substantially larger than what has been found for Bi$_2$Te$_3$\cite{sofote} and comparable to $\tau$ in simple metals. Keeping $\tau$ constant for all doping levels, we calculate the power factor as a function of doping, Fig.~\ref{fig:Power}a. It can be seen that \CBT\ has a high power factor comparable to that of Bi$_2$Te$_3$ and that the estimated experimental carrier concentration is close to optimal for a $p$-doped sample. Fig.~\ref{fig:Power}b shows the calculated $zT$ using the experimental lattice thermal conductivity, $\kappa_l=0.6$~W/mK. 

Fig.~\ref{fig:Power}a also reveals that the optimal power factor and $zT$ for $n$-doped \CBT\ is even higher than for optimally $p$-doped. The high estimated power factor and $zT$ for $n$-doped \CBT\ is interesting considering that $n$-doped \CBT\ samples have been successfully synthesized\cite{CsBiTe,CsBiTe2} and a thermoelectric device needs both $p$- and $n$-type legs. Experimentally the $n$-type compounds are found to have a smaller absolute Seebeck coefficient and a higher conductivity. This can be expected if the carrier concentration is higher than in $p$-type compounds. However, this explanation disagrees with the observation that the Seebeck coefficient starts to decrease at a lower temperature for the $n$-type than for the $p$-type. This can only be explained if the $n$-type materials have a larger concentration of impurities, giving rise to states in the band gap. Furthermore it should be pointed out that electrons and holes couple differently to the lattice vibrations. At 250~K the scattering mechanisms are dominated by electron-phonon coupling and one should use different relaxation times for electrons and holes. Unfortunately the poor agreement between the calculated and experimental temperature dependence of $S$ makes it difficult to estimate a carrier concentration for the $n$-doped samples. The results for $n$-doped materials should therefore be taken with some caution but the results do suggest that further attempts at optimizing the doping for $n$-type materials could be worthwhile.

\subsection{Bonding and band structure}
\begin{figure}
\includegraphics[width=0.5\linewidth]{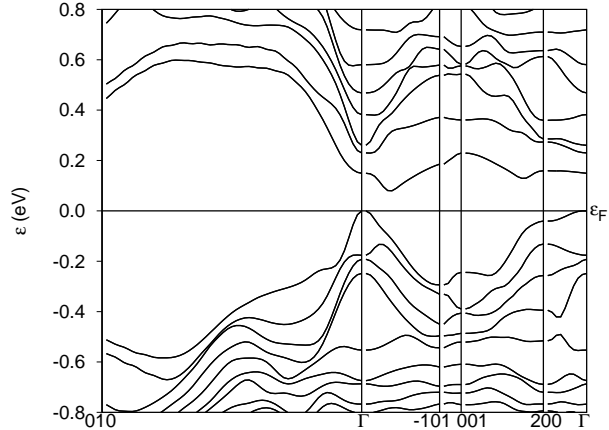}
\caption{Band structure of \CBT. The points are labeled with respect to $\pi(a^*,b^*,c^*)$ where $a^*,b^*$ and $c^*$ are the reciprocal lattice vectors of the  $C$-centered unit cell (see Fig.~\ref{fig:FS}). The band gap fixed at 0.08~eV.}
\label{fig:band}
\end{figure}
Fig.~\ref{fig:band} shows the band structure of \CBT\ with $E_g=0.08$~eV. It is seen that the maximum of the HVB is found at the $\Gamma$-point while the minima of the LCB lie in the $a^*c^*$ plane at the (0.259,0,-0.166) and (-0.259,0,0.166) points. With the indirect gap set to 0.08~eV, a direct band gap of 0.1~eV is found, in good agreement with the optical absorption measurements which also found a gap around 0.1~eV.\cite{CsBiTe2}

\begin{figure*}
\caption{Constant energy surfaces: (a) 0.05~eV below the optimal chemical potential of the HVB (see Fig.~\ref{fig:band}). (b) 0.05~eV above the optimal chemical potential of the LCB (see Fig.~\ref{fig:band}). The lattice vectors of the reciprocal $C$-centered unit cell are shown.} 
\label{fig:FS}
\end{figure*}

It can be shown by a simple argument, that the power factor grows with increasing slope and decreasing height of the transport distribution\cite{sofote} and good thermoelectric properties can be related to a large anisotropy in the transport distribution around the chemical potential. This is caused by the anti-symmetric shape of $(\varepsilon-\mu)\partial f/\partial\varepsilon$, Fig.~\ref{fig:dfde}, that enters the expression for the Seebeck coefficient. The optimal carrier concentration will therefore correspond to a chemical potential close to the band edges. The extrema of the $(\varepsilon-\mu)\partial f/\partial\varepsilon$ function at 250~K are located approximately at $|\varepsilon-\mu|=0.05$~eV, Fig.~\ref{fig:dfde}. From the shape of $(\varepsilon-\mu)\partial f/\partial\varepsilon$ two favorable features for thermoelectric performance can be inferred: (i) The constant energy surface approximately 0.05~eV from the band edge should have a large area which decreases rapidly when approaching the band edge and (ii) the carriers at these energies should have reasonably high group velocities to maximize the transport distribution, Eqs.~(\ref{eq:transdist}-\ref{eq:sigxx}). Normally, the area of the Fermi surface is inversely proportional to the group velocity, so the two conditions above can only be achieved if there is a large anisotropy in the band structure or several carrier pockets contributing to the conductivity.

From Fig.~\ref{fig:band} it is clear that the large surface area is due to the small dispersion of the LCB in the $a^*c^*$-plane, especially along the $a^*$ axis, and the HVB along the $a^*$ direction. This is also illustrated in Fig.~\ref{fig:FS}, which shows the constant energy surfaces at 0.05~eV below and above the HVB and LCB edges. These low dispersion directions co-exist with large dispersions along the $b^*$ axis leading to a good conductivity parallel to the $b$-axis in direct space.

The chemical question is then obviously why the dispersion along the $a^*$-axis is small. To analyze this the constant electron density surface was calculated from the HVB state at the $\Gamma$-point, Fig.~\ref{fig:bond}. As was also pointed out earlier\cite{CsBiTe_larson} the density at this state is mainly situated in sheets along the $a$-axes and it turns out that the density only changes very little for the HVB state at $a^*$, resulting in the small energy dispersion. As opposed to what was reported earlier, also the minimum of the LCB is associated with the same atoms, but now with a clear anti-bonding character.
 
\begin{figure*}
\bmpt{.05}
(a)
\emp
\bmpt{.44}
\emp
\bmpt{.05}
(b)
\emp
\bmpt{.44}
\emp
\caption{(color online) Constant electron density surfaces calculated from (a) the highest valence band state at the $\Gamma$-point and (b) the lowest conduction band at the band edge (the (0.260,0,-0.166) point, see text). The surface covers a volume where the electron density is larger than 0.005 e/\AA. The small spheres in the middle are the Cs, the dark grey/blue sphere represent Bi and the light grey/green spheres represent Te.}  
\label{fig:bond}
\end{figure*}

\section{Conclusion}
We have calculated the electronic structure of the promising thermoelectric compound \CBT\ and shown that a band gap of 0.08~eV must be enforced to obtain a quantitative agreement between the calculated and measured transport properties. It was also shown that a gap in reasonable agreement with experiment can be obtained using the EV-GGA functional. We have found that the experimental $p$-type sample has a carrier concentration close to optimal. The conduction bands have a form equally well suited for thermoelectric properties and we predict that the $n$-type compound might have thermoelectric properties exceeding those of the $p$-type.

\section{Acknowledgments}
GKHM thanks the Carlsberg foundation for financial support. The use of the
computer facilities at the Danish Center for Scientific Computing (DCSC),
Odense, Denmark is acknowledged. 

\end{document}